\documentclass[aps,preprint]{revtex4}%
\usepackage{amsfonts}
\usepackage{amsmath}
\usepackage{amssymb}
\usepackage{graphicx}%
\newtheorem{theorem}{Theorem}

\newtheorem{lemma}{Lemma}

\newenvironment{proof}[1][Proof]{\noindent\textbf{#1.} }{\ \rule{0.5em}{0.5em}}
\begin{document}
\preprint{Internal Communication }
\title{The CMB Dipole and Circular Galaxy Distribution}
\author{Yukio Tomozawa}
\affiliation{Michigan Center for Theoretical Physics and}
\affiliation{Randall Laboratory of Physics}
\affiliation{University of Michigan}
\affiliation{Ann Arbor, MI. 48109-1040 USA}
\date{\today }

\begin{abstract}
The validity of Hubble's law defies the determination of the center of the big
bang expansion, even if it exists. Every point in the expanding universe looks
like the center from which the rest of the universe flies away. In this
article, the author shows that the distribution of apparently circular
galaxies is not uniform in the sky and that there exists a special direction
in the universe in our neighborhood. The data is consistent with the
assumption that the tidal force due to the mass distribution around the
universe center causes the deformation of galactic shapes depending on its
orientation and location relative to the center and our galaxy. Moreover, the
cmb dipole data can also be associated with the center of the universe
expansion, if the cmb dipole at the center of our supercluster is assumed to
be due to Hubble flow. The location of the center is estimated from the cmb
dipole data. The direction to the center from both sets of data is consistent
and the distance to the center is computed from the cmb dipole data.

\end{abstract}

\maketitle

\section{\label{sec:level1}Introduction}

Since the discovery of Hubble's law in 1929 and the big bang interpretation of
its data, the question lingers whether a center for the expansion exists and
if so where. The Hubble law, \textbf{v }= H$_{\text{0}}$ \textbf{r}, yields
the relationship, \textbf{v}$_{\text{2}}$ - \textbf{v}$_{\text{1}}$=
H$_{\text{0}}$ ( \textbf{r}$_{\text{2}}$ - \textbf{r}$_{\text{1}}$) for any
two galaxies with positions and velocities, \textbf{r}$_{\text{1}}$,
\textbf{v}$_{\text{1}}$ and \textbf{r}$_{\text{2}}$ , \textbf{v}$_{\text{2}}$
respectively, where H$_{\text{0}}$ = 100 h km/s Mpc is the Hubble constant
(with h = 0.5 \symbol{126}0.85). For convenience of discussion, we assume the
value of H$_{\text{0}}$ to be 70.0 km/s Mpc in this article. The last equation
implies that every point appears to be the center of the expansion. Besides,
the distribution of galaxies surrounding us seems to be isotropic on some
large distance scale. In order to get information about the existence of a
center for expansion of the universe, one has to search for observational data
beyond the Hubble law. The author will show that such information can be
obtained from data for the distribution of circular galixies and the cmb
dipole in the temperature distribution. These independent data indicate that
the direction to the center is consistent for both, and one set of data gives
a determination of the distance.

\section{Distribution of Circular Galaxies}

Circular galaxies appear when the circular faces of spiral galaxies or
elliptical galaxies are facing us; otherwise they are elliptical in shape. The
fraction of circular galaxies in any direction depends on the probability of
the orientation of the symmetry axis coinciding with the line of sight. If one
assumes that such a probability is uniform in any direction, then one expects
the distribution of circular galaxies to be isotropic. The author will show
that that is not the case for the existing data.

Fig. 1 (a) shows the distribution of 23,011 bright galaxies in galactic
coordinates compiled in $\ $RC3\cite{rc3}. The galaxies in the galactic plane
are missing for the obvious reason that they are prevented from observation
due to our own Milky Way. Plots in equatorial coordinates and supergalactic
coordinates show a similar uniform distribution. The number of galaxies on the
two sides of the galactic plane is 12,323 in the north and 10,628 in the south
with a ratio of 1.1651. The difference in these numbers may be partly due to a
statistical fluctuation and partly due to a historical bias in past
observation. Anyway, any study hereafter can be normalized with this ratio.
Fig. 1 (b) shows the velocity distribution for the compiled galaxies. The
number of galaxies for which the velocities, cz, are known is 10633 (5923 in
the north and 4710 in the south). The approximate upper bound is $\sim$15,000
km/s and the average, $\sim$7,500 km/s, corresponds to a distance of 110 Mpc.

In Fig. 2 (a), (b) and (c), we show the distribution of galaxies with circular
shape in galactic, supergalactic and equatorial J2000.0 coordinates,
respectively. These galaxies correspond to those with the value R25 = 0.00
(the error is in the range of $\sim$0.10) in RC3, where R25 stands for the
logarithm of the ratio of the apparent major and minor axes. The number of
such galaxies is 682 in the north, 416 in the south with 1098 for the total.
Also shown in Fig. 2 (d) is the velocity distribution, which is similar to Fig
1 (b). The pertinent characteristics of Fig. 2 are the following:

(i) Apparently circular galaxies are not distributed uniformly in the sky. If
the orientation of galaxies is distributed at random, Fig 2 (a)-(c) should
show statistically uniform distributions. It definitely shows directionality.

(ii) The southern part is more compact compared to the northern part. Since a
closed curve surrounding the pole is extended to all longitudes in Fig 1 (a),
the views in other coordinates, supergalactic coordinates in Fig 2 (b) and
equatorial coordinates J2000.0 in Fig 2 (c), provide intuitively clearer
distribution plots. Fig 2 (b) shows similar shapes for both, northern and
southern, distributions. The normalized ratio of the northern to the southern
distribution is (682)/(416)/(1.1651) = 1.4071. The shapes in supergalactic
coordinates in Fig 2 (b) match this ratio. Note that the left and right blobs
in Fig 2 (b) approximately correspond to those in the north and the south in
Fig 2 (a).

Fig 3 (a)-(d) shows the distribution of galaxies with R25 = 0.01-0.10,
0.11-0.20, 0.21-0.40 and 0.41-1.00. They do not show a distinct directionality
except for a higher density in the neighborhood of (120d, -30d). The total
number of galaxies, the numbers in the north and south as well as the
normalized north/south ratio are given for sliced values of R25 in Table~1. In
summary, a distinct directionality exists only for R25 = 0.00, and not for the
other values of R25.

The meaning of this directionality in the distribution of apparently circular
galaxies is an important question .The author suggests that the existence of a
center for universe expansion is consistent with the presented data and
suggests the direction to the center, if not the determination of the distance.

(I) Galaxies are subject to tidal forces due to the gravity of the masses
which are contained in a sphere of radius equal to the distance from the
center to the galaxies in question. Circular galaxies in the radial direction
from the center through us stay circular in the presence of tidal forces due
to axial symmetry, while those in directions perpendicular to the radial seen
from us are deformed by the squeeze and stretch of tidal forces and therefore
become elliptical. This explains the directionality in Fig. 2 (a)-(c).

(II) If the center lies at a large distance compared with the observed
galaxies in the RC3 catalog, then the circular galaxy blobs in the both
hemispheres should be approximately equal in size, since the defining cone for
circular invariance constitutes parallel lines. The disparity in the south and
the north suggests that the center is nearby and the direction to the center
must be towards the north, since the tidal force near the center is small so
that the distortation is minimal in the neighborhood. This explains the larger
extension of circular galaxies in the northern hemisphere. The central value
of the direction of the center is in the neighborhood of
\begin{equation}
l_{c}=120%
{{}^\circ}%
+180%
{{}^\circ}%
=300%
{{}^\circ}%
\pm15%
{{}^\circ}%
,\text{ \ \ \ }b_{c}=30%
{{}^\circ}%
\pm15%
{{}^\circ}%
. \label{center}%
\end{equation}
The distance to the center, $R_{c}$, is bounded by 10000km/s ==%
$>$%
143Mpc.

(III) If the matter density is constant, it produces a gravitational force
that is linear in the radial direction, thus yielding a tidal force that is
spherically symmetric and attractive. Obviously, such a tidal force does not
explain the data of Fig.2 (a)-(c). Although the assumption of homogeneity and
isotropy in the Friedman-Robertson-Walker metric yields a constant matter
density at a fixed time, experimental evidence for those assumptions is yet to
come. As a matter of fact, the analysis of this article finds that the matter
density of the universe is not constant on this length scale. We are
discussing a length scale that involves our local supercluster which is not
uniform in density.

In Fig 4(a), circular galaxy distribution (R25 = 0) from a newer compilation,
the Hyperleda catalog (PGC 2003) with 12,008 items out of a total of 983,261
is shown. Three blobs in the equatorial slab are stars. Away from the
equatorial slab, most of data points are galaxies. The data away from the
equatorial slab has features similar to that of the data from the RC3 catalog,
Fig 2(a). Fig 4(b) shows a spherical plot of Fig 4(a).

\section{The cmb Dipole in the Temperature Distribution}

A dipole component was observed in the temperature distribution in the cmb
(cosmic microwave background radiation) measurement. The blue shift dipole
value for the solar system \cite{dipole} is v = 371$\pm$0.5 km/s, l =
264.4$\pm$0.3$%
{{}^\circ}%
$, b = 48.4$\pm$0.5$%
{{}^\circ}%
$. Or alternatively, the redshift dipole value is%
\begin{equation}
v=371\pm0.5\text{ }km/s,\text{ \ \ \ }l=84.4\pm0.3%
{{}^\circ}%
,\text{ \ \ \ }b=-48.4\pm0.5%
{{}^\circ}%
.
\end{equation}
By using the observation of a peculiar velocity for the solar system, one can
compute the dipole component of the cluster (Virgo) center and that of the
supercluster (Great Attractor) center. Then, one can associate the last dipole
component with Hubble flow. In order to understand the relationship among
Hubble flow, peculiar velocity and the cmb dipole, the author presents
pertinent theorems.

\begin{theorem}
With the assumption of the existence of a center for expansion of the
universe, Hubble flow creates a cmb dipole with red shift in the direction of
the Hubble flow with the magnitude of the Hubble flow velocity.
\end{theorem}

\begin{proof}
Let the velocities of the Hubble flow and the cmb emitter in the direction of
the Hubble flow be $v_{H}$ and $v$, respectively. The relative velocity of the
two is $v_{+}=(v-v_{H})/(1-vv_{H}/c^{2})$. The analogous velocities in the
opposite direction are $-v_{H}$ and $v-2v_{H}$, and the relative velocity is
$v_{-}=(v-2v_{H}+v_{H})/(1+(v-2v_{H})v_{H}/c^{2})$ in the opposite direction.
The difference of these opposite relative velocities is then
\begin{equation}
v_{+}-v_{-}=(v-v_{H})(2v-2v_{H})v_{H}/c^{2}(1-vv_{H}/c^{2})(1+(v-v_{H}%
)v_{H}/c^{2})\approx2v_{H},\text{ \ \ \ \ \ \ \ }for\text{ \ }v\approx c\text{
and }v_{H}\ll c.
\end{equation}
and%
\begin{equation}
v_{\pm}\approx v-v_{H}\pm v_{H}%
\end{equation}
This proves the statement. The cmb dipole vanishes at the center of the universe.
\end{proof}

\begin{theorem}
The cmb dipoles for red shift at the center of a cluster and at a member
galaxy with a peculiar velocity $\mathbf{v}_{p}$ are related by%
\begin{equation}
\mathbf{v}(dipole\text{ }at\text{ }the\text{ }clustercenter)=\mathbf{v}%
(dipole\text{ }at\text{ }the\text{ }galaxy)-\mathbf{v}_{p}+\mathbf{v}%
(clustercenter-galaxy),
\end{equation}
where%
\begin{equation}
\mathbf{v}(A-B)=\mathbf{v}(A)-\mathbf{v}(B).
\end{equation}

\begin{proof}
The peculiar velocity plays a role similar to Hubble flow, as far as the cmb
dipole is concerned. This can be seen by considering the case where a peculiar
velocity, $\mathbf{v}_{p}$, and a Hubble flow, $\mathbf{v}_{H}$, are parallel
to each other. In this case, the galaxy is at rest with respect to a galaxy
with a Hubble flow, $\mathbf{v}_{H}+\mathbf{v}_{p}$. Since both galaxies
should have the same cmb dipole, a peculiar velocity yields a redshift cmb
dipole with $\mathbf{v}_{p}$. That is the reason for the subtraction. The
term, $\mathbf{v}(clustercenter-galaxy)=$ $\mathbf{v}%
(clustercenter)-\mathbf{v}(galaxy)$, is for adjustament of the Hubble law.
\end{proof}
\end{theorem}

\begin{lemma}
The center of the universe should have an intrinsically vanishing value for
the cmb dipole.
\end{lemma}

Hereafter a cmb dipole implies a redshift dipole unless otherwise stated. The
peculiar velocity of the sun relative to the Virgo center of the local cluster
is estimated to be \cite{sciama}
\begin{equation}
v=415\text{ }km/s,\text{ \ \ \ }l=335%
{{}^\circ}%
,\text{ \ \ \ }b=7%
{{}^\circ}
\label{sciama1}%
\end{equation}
or%
\begin{equation}
v=630\text{ }km/s,\text{ \ \ \ }l=330%
{{}^\circ}%
,\text{ \ \ \ }b=45%
{{}^\circ}
\label{sciama2}%
\end{equation}
We examine these two possibilities. The outcomes are listed in this order for
each case. Using Theorem 2, one computes the cmb dipole at the Virgo center,
which is located at%
\begin{equation}
v=1050\pm200\text{ }km/s,\text{ \ \ \ }l=287%
{{}^\circ}%
,\text{ \ \ \ }b=72.3%
{{}^\circ}%
\end{equation}
corresponding to a distance of $15\pm3$ $Mpc$. Hereafter, all distances are
expressed in the form of receding velocities due to the Hubble law,
$\mathbf{v}=H_{0}\mathbf{r}$.\ The key formula is%
\begin{equation}
\mathbf{v}(dipole\text{ }at\text{ }Virgo)=\mathbf{v}(dipole\text{ }at\text{
}the\text{ }Sun)-\mathbf{v}_{p}(Sun/Virgo)+\mathbf{v}(Virgo). \label{dipole1}%
\end{equation}
For the cmb dipole at the Virgo center, one obtains
\begin{equation}
v=728.3\pm148\text{ }km/s,\text{ \ \ \ }l=336.0\pm8.2%
{{}^\circ}%
,\text{ \ \ \ }b=67.4\pm11.3%
{{}^\circ}%
\end{equation}%
\begin{equation}
v=418.8\pm47\text{ }km/s,\text{ \ \ \ }l=328.8\pm6.4%
{{}^\circ}%
,\text{ \ \ \ }b=41.5\pm28.0%
{{}^\circ}%
\end{equation}
depending on the two choices of peculiar velocity. We note that the Cartesian
coordinates for $(v,l,b)$ are expressed as $(v\cos(b)\cos l,v\cos(b)\sin
l,v\sin(b).$

Further, the Virgo cluster is considered to be a part of a supercluster
centered around the GA (great attractor). In order to compute the cmb dipole
at the GA, one assumes the velocity of the Virgo center to be corresponding to
a distance of $15\pm3$ $Mpc$. The position of the GA is taken to be \cite{ga}%
\begin{equation}
v=4200\text{ }km/s,\text{ \ \ \ }l=309%
{{}^\circ}%
,\text{ \ \ \ }b=18%
{{}^\circ}
\label{eqga1}%
\end{equation}
or%
\begin{equation}
v=3000\text{ }km/s,\text{ \ \ \ }l=305%
{{}^\circ}%
,\text{ \ \ \ }b=18%
{{}^\circ}
\label{eqga2}%
\end{equation}
and the infall velocity of the Virgo center to the GA to be
\begin{equation}
v_{in}=1000\pm200km/s.
\end{equation}
The direction of the infall is determined by%
\begin{equation}
\mathbf{v}(GA-V)=\mathbf{v}(GA)-\mathbf{v}(Virgo)
\end{equation}
resulting in%
\begin{equation}
v(GA-V)=3712.3\pm75\text{ }km/s,\text{ \ \ \ }l(GA-V)=310.9\pm0.4%
{{}^\circ}%
,\text{ \ \ \ }b(GA-V)=4.6\pm2.8%
{{}^\circ}%
\text{ \ \ \ } \label{infall1}%
\end{equation}
for Eq.(\ref{eqga1}) and%
\begin{equation}
v(GA-V)=2552.5\pm59\text{ }km/s,\text{ \ \ \ }l(GA-V)=307.2\pm1.6%
{{}^\circ}%
,\text{ \ \ \ }b(GA-V)=1.6\pm3.0%
{{}^\circ}
\label{infall2}%
\end{equation}
for Eq. (\ref{eqga2}).

Using Theorem 2, one can compute the cmb dipole at the GA by%
\begin{equation}
\mathbf{v}(dipole\text{ }at\text{ }GA)=\mathbf{v}(dipole\text{ }at\text{
}Vigo)-\mathbf{v}(Virgo\text{ infall})+\mathbf{v}(GA-V), \label{dipole2}%
\end{equation}
where the direction of the infall is given by Eq. (\ref{infall1}) or Eq.
(\ref{infall2}). Then, the cmb dipole at the GA is given by%
\begin{equation}
v=2609.3\pm120\text{ }km/s,\text{ \ \ \ }l=308.1\pm0.2%
{{}^\circ}%
,\text{ \ \ \ }b=19.9\pm3.4%
{{}^\circ}%
\end{equation}%
\begin{equation}
v=2459.1\pm98\text{ }km/s,\text{ \ \ \ }l=308.6\pm0.6%
{{}^\circ}%
,\text{ \ \ \ }b=11.6\pm3.9%
{{}^\circ}%
\end{equation}
for Eq. (\ref{eqga1}), and%
\begin{equation}
v=1455.6\pm142\text{ }km/s,\text{ \ \ \ }l=301.3\pm0.6%
{{}^\circ}%
,\text{ \ \ \ }b=25.5\pm5.3%
{{}^\circ}%
\end{equation}%
\begin{equation}
v=1286.6\pm104\text{ }km/s,\text{ \ \ \ }l=302.0\pm0.7%
{{}^\circ}%
,\text{ \ \ \ }b=10.4\pm7.3%
{{}^\circ}%
\end{equation}
for Eq. (\ref{eqga2}).

Based on Theorem 1, one may assume that the cmb dipole at the GA calculated
above is due to the Hubble flow of the the center of the GA. Then, one can
compute the position of the center for expansion of the universe by%
\begin{equation}
\mathbf{v}(universe\text{ }center)=\mathbf{v}(GA\text{ }center)-\mathbf{v}%
(the\text{ }cmb\text{ }dipole\text{ }at\text{ }GA) \label{dipole3}%
\end{equation}
The position of the Universe center thus obtained is%
\begin{equation}
v_{c}=1595.3\pm196\text{ }km/s,\text{ \ \ \ }l_{c}=310.5\pm0.1%
{{}^\circ}%
,\text{ \ \ \ }b_{c}=14.8\pm1.5%
{{}^\circ}
\label{center1}%
\end{equation}%
\begin{equation}
v_{c}=1779.6\pm184km/s,\text{ \ \ \ }l_{c}=309.7\pm0.2%
{{}^\circ}%
,\text{ \ \ \ }b_{c}=26.8\pm2.7%
{{}^\circ}
\label{center2}%
\end{equation}
for Eq. (\ref{eqga1}), and%
\begin{equation}
v_{c}=1585.4\pm196\text{ }km/s,\text{ \ \ \ }l_{c}=308.1\pm0.2%
{{}^\circ}%
,\text{ \ \ \ }b_{c}=13.0\pm1.6%
{{}^\circ}
\label{center3}%
\end{equation}%
\begin{equation}
v_{c}=1759.3\pm181\text{ }km/s,\text{ \ \ \ }l_{c}=307.4\pm0.0%
{{}^\circ}%
,\text{ \ \ \ }b_{c}=25.3\pm2.9%
{{}^\circ}
\label{center4}%
\end{equation}
for Eq. (\ref{eqga2}). Conversion to the ordinary distance scale yields
$22.8\pm2.8$ $Mpc$, $25.4\pm2.6$ $Mpc$ for Eq. (\ref{eqga1}) and $22.6\pm2.8$
$Mpc$, $25.1\pm2.6Mpc$ for Eq. (\ref{eqga2}). The angles obtained here for the
direction of the universe center are quite consistent with those obtained in
the previous section from the circular galaxy distribution, Eq. (\ref{center}).

Combining all the processes, Eq. (\ref{dipole1}), Eq.(\ref{dipole2}) and Eq.
(\ref{dipole3}), one gets%
\begin{equation}
\mathbf{v}(universe\text{ }center)=-(\mathbf{v}(dipole\text{ }at\text{
}the\text{ }Sun)-\mathbf{v}_{p}(total))
\end{equation}
where%
\begin{equation}
\mathbf{v}_{p}(total)=\mathbf{v}_{p}(Sun/Virgo)+\mathbf{v}_{p}(Virgo/GA)
\end{equation}
is the total peculiar velocity of the sun towards the GA. This equation gives
identical results for the center of the universe, Eq. (\ref{center1})-Eq.
(\ref{center4}). In other words, the author has arrived at the following theorem.

\begin{theorem}
The Hubble flow of the solar system is given by%
\begin{equation}
v_{H}(the\text{ }Sun)=\mathbf{v}(the\text{ }cmb\text{ }dipole\text{ }at\text{
}the\text{ }Sun)-\mathbf{v}_{p}(the\text{ }total\text{ peculiar }%
velocity\text{ }of\text{ }the\text{ }Sun)
\end{equation}

\end{theorem}

The similarity of the final results for the location of the center of the
universe, Eq. (\ref{center1})-Eq. (\ref{center2}) and Eq. (\ref{center3})-Eq.
(\ref{center4}), is easily understood from this theorm, since the directions
of the two solutions for the GA position, Eq. (\ref{eqga1}) and Eq.
(\ref{eqga2}), are similar and therefore the peculiar velocities from Virgo
towards the GA are similar, despite a large difference between the distances
to the GA. Therefore, the difference comes from the peculiar velocity for the
sun towaeds the Virgo center, Eq. (\ref{sciama1}) or Eq. (\ref{sciama2}). In
order to settle on the distance to the universe center, one has to settle on
one of the two choices for the peculiar velocity, Eq. (\ref{sciama1}) or Eq.
(\ref{sciama2}).

\section{Summary and Discussion}

The data compiled in RC3 is an accumulation by various groups, so that the
criterion for circularity may not be uniform across observational groups.
Nevertheless, the consistency between the cmb dipole data and the circular
galaxy distribution for the direction of the universe center suggests
consistency among the many groups that contributed to the RC3 catalog in that
respect. The recent measurement by SDSS increases the data on galaxies in
number and quality. It is desirable, however, to enlarge the scope of sky
coverage into the southern hemisphere to get improved data for the circular
galaxy distribution. Finally, the author suggests that the deformation of
galaxies by cosmic tidal forces discussed in this article be taken into
account for the systematic study of the triaxiality of galaxies \cite{triax}
and asymmetric spiral galaxies \cite{spiral}. From statistical analysis of
these galaxy deformations, one may be able to reach a more accurate
determination of the new cosmological parameter, the position of the center of
the universe.

As for the discussion of the cmb dipole, the observed cmd dipole cannot be
explained merely by the presence of the peculiar velocity of the solar system.
One should be involved with the Hubble flow even only from dimensional
considerations. Theorem 3 suggests an important departure from normal thinking
in cosmology. If further information on the peculiar velocity of the GA
supercluster is found in the future, a similar procedure should be applied to
the new situation. Then, the final result for the center of the universe
expansion might be changed. However, the conclusion from the circular galaxy
distribution gives a stringent constraint on the distance and the direction of
the universe center.

There is a proposal and discussion of a rotation of the universe based on data
of polarization of radio waves\cite{birch}. In general, an intrinsic
polarization of radio waves can be related to the existence of the center of
the universe, since the deformation of a galaxy by tidal forces can cause an
asymmetric magnetic field that creates polarization at the source which is
called intrinsic. Then, some features of the observed polarization can be
explained by the mere existence of the center of the universe, but it is not
clear whether all the features can be explained from that or not. If not, one
can ask the question whether there exists a unified description of all the
data by the center and a rotation of the universe. Certainly, any rotation
axis of the universe should pass through the center in a simple picture. In
fact, the angles reported in the two works are close to each other: For the
center of the universe, (l, b) = (310$%
{{}^\circ}%
$, 30$%
{{}^\circ}%
$) --%
$>$
(ra, dec) = (13h17m, -32d), and for the rotation axis, (ra, dec) =
(14h35m,-35d) with an error bar of (30m, 30m). The proximity of these angles
would make the construction of a unified picture easier.

\bigskip

\begin{acknowledgments}
The author would like to thank the members of the Physics Department and the
Astronomy Department of the University of Michigan for useful information.
\end{acknowledgments}

Correspondence should be addressed to the author at tomozawa@umich.edu.

\bigskip

Figure Caption.

Fig.1 Distribution of bright galaxies compiled in RC3 \cite{rc3}. (a) Galaxy
distribution in galactic coordinates. (b) Velocity distribution vs. galactic latitudes.

Fig.2 The distribution of apparently circular galaxies which have the value
R25= 0.00: (a) In galactic coordinates. (b) In supergalactic coordinates. (c)
In equatorial J2000.0 coordinates. (d) Velocity distribution vs. galactic latitudes.

Fig.3 Distribution of galaxies in various ranges of the value of R25: (a)
0.01-0.10, (b) 0.11-0.20, (c) 0.21-0.40 and (d) 0.41-1.00.

Fig.4(a): Circular galaxy distribution (R25 = 0) from the Hyperleda catalog
(PGC 2003) in galactic coordinates (12,008 items out of a total of 983,261).

Fig.4(b): Spherical plot of Fig.4(a).

\bigskip

Table Caption.

Table1. The total number of galaxies, the numbers in the north and south and
the normalized north/south ratio in various ranges of R25.

\bigskip

\bigskip%
\begin{tabular}
[c]{lllll}%
Range of R25 & Total & North & South & North/South Normalized Ratio\\
0.00-0.00 & 1098 & 682 & 416 & 1.4071\\
0.01-0.10 & 4573 & 2330 & 2243 & 0.8916\\
0.11-0.20 & 4448 & 2322 & 2126 & 0.9374\\
0.21-0.30 & 3286 & 1714 & 1572 & 0.9358\\
0.31-0.40 & 2246 & 1148 & 1098 & 0.8974\\
0.41-0.50 & 1638 & 861 & 787 & 0.9281\\
0.51-1.00 & 4050 & 2210 & 1835 & 1.0337
\end{tabular}

\bigskip
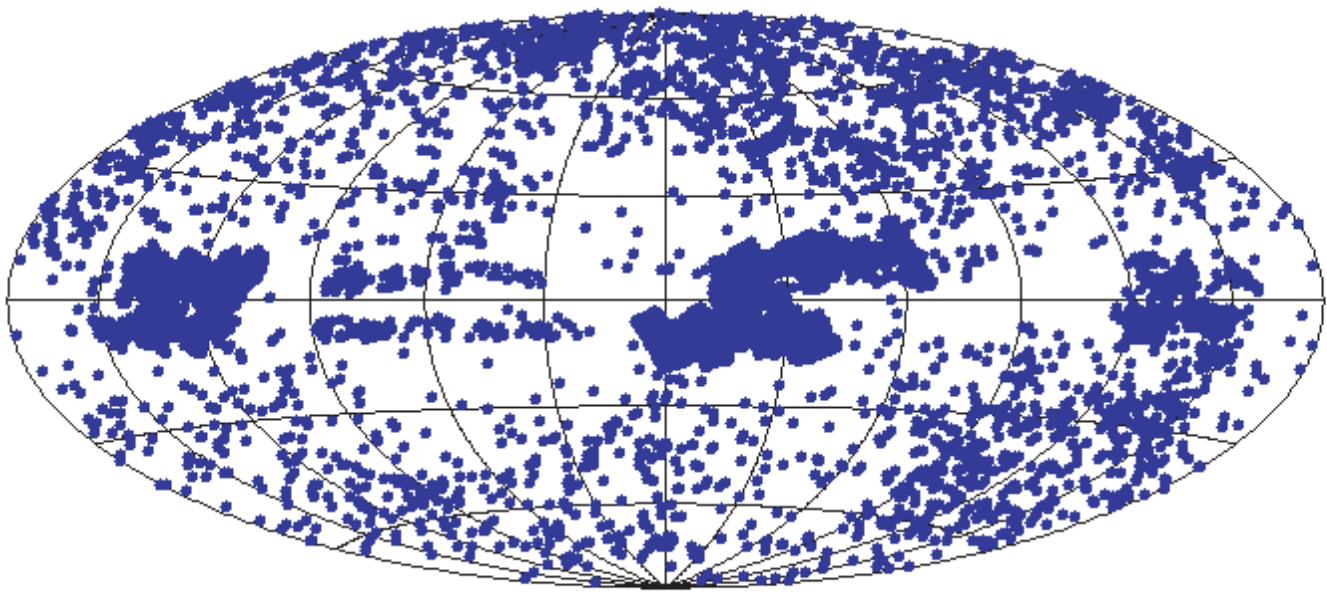
\end{document}